# Fabrication of graphene nanogap with crystallographically matching edges and its electron emission properties

H. M. Wang,[1] Z. Zheng,[1] Y.Y.Wang,[1] J.J. Qiu,[2] Z.B. Guo,[2] Z. X. Shen,[1] and T. Yu [1,*]

1 Division of Physics and Applied Physics, School of Physical and Mathematical Sciences, Nanyang Technological University, 1 Nanyang Walk, Block 5, Level 3, Singapore 637616

2 Data Storage Institute, 5 Engineering Drive 1, Singapore 117608

We demonstrate the fabrication of graphene nanogap with crystallographically matching edges on $SiO_2$/Si substrates by divulsion. The current-voltage measurement is then performed in a high-vacuum chamber for a graphene nanogap with few hundred nanometers separation. The parallel edges help to build uniform electrical field and allow us to perform electron emission study on individual graphene. It was found that current-voltage (I-V) characteristics are governed by the space-charge-limited flow of current at low biases while the Fowler–Nordheim model fits the I-V curves in high voltage regime. We also examined electrostatic gating effect of the vacuum electronic device. Graphene nanogap with atomically parallel edges may open up opportunities for both fundamental and applied research of vacuum nanoelectronics.

* Electronic mail: yuting@ntu.edu.sg

Since the discovery of large size graphene in 2004,[1] it inspired researchers due to its unique electronic properties, including linear energy dispersion relationship,[2] high carrier mobility,[3] ballistic transport,[4] and quantum interference[5,6]. The recent experiments of width[7] and layer[8] engineering have demonstrated the potential applications of graphene in nanoeletronics. In addition, graphene may be of the potential in the application of vacuum electronics, like flat-panel displays, and microwave amplifiers for two reasons: (1) Graphene has atomically sharp edges with the thickness of 0.34nm. The nature of sharpness is supposed to give a great enhancement of field strength as an emitter.[9] Electrons can be extracted from the edges by tunneling through the surface potential barrier under a comparatively small bias; (2) Its metallic nature and low contact resistance[10] with metals give a small voltage drop both along the graphene and at the graphene/metal interface. This means that there are no factors significantly obstructing the supply of electrons to the emitter. Recently, the electric field emission behavior of vertically aligned few-layer graphene was studied in a parallel plate–type setup.[11,12,13] The low turn-on electrical field,[11] high emission current density,[11] high emission stability,[12] and long lifetime,[12] make graphene sheets an efficient emitter for display backlight sources. However, graphene sheets in previous experiments are in a dense network structure. Electron emission from individual graphene sheet has yet to be reported. In this letter, we will present the fabrication of graphene nanogaps and experimental study on its electron emission properties. The effect of electrostatic gating is also examined.

Graphene nanogaps were fabricated by divulsion with the help of a standard electronic beam lithography process. A graphene layer on top of Polymethyl methacrylate (PMMA) allows us to produce a graphene nanogap on top of a Si/SiO$_2$ substrate in a relatively easy way. Using the model of visibility study,[14] a 100nm thick PMMA layer on top of a 300nm thick SiO$_2$ is found to achieve the highest contrast (8%) centered at the wavelength of 500nm. The fabrication processes are illustrated in Fig. 1. After a 100nm thick PMMA layer was spanned onto SiO$_2$ substrate, graphene flakes were transferred onto the PMMA layer by mechanical exfoliation[1]. After localizing graphene flakes, another 300nm PMMA was coated onto the substrate. Subsequently, electrical contacts Ti(10nm)/Au(130nm) were formed by using a combination of electron beam lithography and evaporation. After liftoff, these contacts became free standing bridge and therefore supported graphene from both sides. The free standing structure of graphene only survived in the solvent. Most of suspended graphene flakes collapsed and were torn apart by the surface tension of solvent during the drying process. The method is one of easy ways to produce a graphene nanogap device with atomically parallel edges.

As proof, optical image, contrast and Raman spectra of single layer graphene (SLG) on top of PMMA(100nm)/SiO$_2$(300nm)/Si are given in Fig. 2a, 2b and 2c, respectively. The 2D peak of the Raman spectrum has a full width at half maximum of 32cm$^{-1}$ which distinguishes a monolayer from few-layer graphene.[15] Scanning electron micrograph (SEM) of a representative SLG nanogap connected with suspended electrodes is shown in Fig. 2d. The length and width of the nanogap are around 2.8μm and 100nm, respectively. In the SEM image, graphene exhibits similar contrast to the substrate, and the brighter edge helps us to profile graphene flakes. The arrows in Fig. 2d point out the edges of graphene nanogap. Before the electrical measurement, the devices were transferred into a vacuum chamber (<10$^{-7}$ Torr) and annealed at 400°C for eight hours to remove amorphous carbon deposited during SEM imaging. The I-V characterization was carried out by Keithley 237 sourcemeter in a vacuum chamber with a base pressure of ~10$^{-6}$ Torr. Another Keithley 237 sourcemeter served to give a gate bias. The across-gap voltage increases from 0V up to 60V, with a ramp-up voltage rise of 1V in each step. The sweeping rate is 1 second/V. A current limit of 50nA is imposed to avoid overheating. The back-gating voltage in the range of -80V up to 80V is applied across SiO$_2$ onto the graphene cathode.

For such a graphene nanogap, understanding of the I-V characteristics has become important for applications in vacuum nanoelectrnics. Experimental I-V curves for the nanogap device (see Fig. 2d) are shown in Fig. 3a. It is found that current increases with voltage, especially for voltages higher than 30V until the compliance of 50nA is approached. We analyzed the I-V curves using the Fowler-Nordheim (F-N) model[16], which is always used to study the field emission process. And then the field enhancement factor β is estimated. According to F-N equation, the emission current is given by:

$$I \propto \frac{F^2}{\Phi} \exp\left(\frac{-B\Phi^{\frac{3}{2}}}{F}\right)$$, where *I* is the current, *B* is a constant given by 6.83×10$^7$ ($V \cdot eV^{-\frac{3}{2}} cm^{-1}$), *Φ*

is work function and $F$ is local electric field. The local electric field is often written as $F = \beta E = \frac{\beta V}{d}$, where $E$ is the applied macroscopic field obtained by an applied voltage $V$ between two graphene electrodes separated with a distance $d$, $\beta$ is field enhancement factor. Assuming that the data in Fig. 3a are described by F-N relation, the corresponding $\ln(\frac{I}{V^2}) \sim \frac{1}{V}$ plots under different gate bias are plotted in Fig. 3b. It is found that the F-N plots yielded a straight line for all the gate biases after the current goes beyond 5 nA and up to the limit 50nA. The results confirm that the current higher than 5nA was mainly resulted from field emission process. By linear fitting, we can obtain a slope which equals to $\frac{B\Phi^{3/2}d}{\beta}$ from the F-N plots, where $d$ is given by the experimental configuration (about 100nm), the work function $\Phi = 5$ eV is used for graphitic materials;[11] therefore, the enhancement factor β can be determined. Fig. 3d shows β extracted from F-N plots as a function of gate voltage. From the figure, it is found that when the gate voltage sweeps from +80V to -80V, the value of β varies around 68 and exhibit very weak dependence on gate voltage. The value of β from the experiments is not as high as several hundred which are expected. There are two possible reasons: 1) As reported in literature,[17] β is proportional to the spacing between anode and the cathode. The small gap spacing between the cathode and anode in the experiments may contribute to the small value of β; 2) The other possible reason is that the field enhancement is limited in the nanogap because of its lateral nature on substrate. Impurities (such as oxygen, water, organic residue) were unavoidably absorbed on the emitter and the substrate when samples were transferred from annealing chamber to FE chamber in atmosphere. The impurities could form dipole and apply an additional disturbance on local electrical field near graphene edge. The disturbance may change the local work-function of the graphene edge.[18] In addition, some electrons emitted from the cathode could be trapped in the impurities in front of cathode; the trapped electrons reduced the local electrical field of the cathode. Therefore, the emission performance could be degraded. The weak gating dependence of β may be due to the fact that the gating bias generates an electrical field at graphene edge rather smaller than the across-gap bias does. The quantitative comparison about the local electrical field near the edge will be given in simulation part.

As shown in Fig. 3b, the electron emission at low bias can not be described by the F-N law. Emission of electron sometimes could become a space-charge limited (SCL) emission, described by the classical Child-Langmuir (CL) law.[19,20,21] Depending on the across-gap voltage ($V$) and gap spacing ($d$), the classical CL law in two-dimensional system is given by $I \propto (\frac{4\varepsilon_0}{9})(\frac{2e}{m_e})^{\frac{1}{2}}(\frac{V^{\frac{3}{2}}}{d^2})$, where $\varepsilon_0$ is the permittivity of free space, $e$ is the elementary charge, and $m_e$ is electron mass. Assuming that the low bias data in Fig. 3a are described by CL relation, the corresponding $\ln(I) \sim \ln(V)$ curves are plotted in Fig. 3c. It is observed that the plots of $\ln(I)$ versus $\ln(V)$ yielded a straight line when $I$ is lower than 5nA. The slope α of the line is extracted from Fig. 3c under different gate voltage. And the relationship of slope α and gate voltage is then plotted in Fig. 3d. It is observed that the slope α of the best fit line fluctuates around 1.5, depending on gate voltage. The value of 1.5 indicates that the classic CL law is effective at the low across-gap bias regime.

Above results reveal that there are two distinct regimes for I-V curves of the graphene nanogap. Our analysis elucidates the transition from CL law to F-N law with the increase of voltage. When a small across-gap bias is applied across two SLG electrodes, the electrons emitted receive scattering by the impurities on $SiO_2$ surface and are trapped to the impurities in front of the cathode. The process suppresses both the electric field near cathodes and the electron emission. As the quantity of impurities is not high, above process saturates after the bias approaches a certain value. Finally, electrons are emitted from the cathode surface following the F-N field emission. In the low across-gap bias regime, the large fluctuation of α with respect to gating biases indicates that the electrical field near SLG edges generated by the gating bias is comparable to that generated by the across-gap bias. The transition was also observed in the other two samples we examined.

In order to understand the experimental results further, we investigated the spatial distribution of electric field inside a graphene nanogap via finite element modeling. The across-gap voltage of 60V was applied across a nanogap with 100nm separation and 100nm width. Note that the dimension of the device model does not match the real sample because of the limitation of our workstation. However, simulation results still make sense in understanding the experimental results of the real nanogap device. According to the simulation results,[22] very high gradients of electric field are found near graphene edges. The field strength in the gap is around $1.35\times10^9$ V/m along the edges of the nanogap. The maximum field strength is about $3.26\times10^9$ V/m at the two corners of SLG edges. Let's turn to the electric field created by electrostatic gating voltage. The gating voltage of 80V applied to the graphene device creates an electrical field of about $3\times10^8$ V/m across the silicon oxide. The magnitude of electrical field is almost one order smaller than that induced by the across-gap bias at graphene edges. Assuming that the gating bias keeps 80V and the across-gap bias decreases to several voltages, the strength of electrical field created by both biases at the edge of graphene becomes comparable. Under this situation, electrons emitted can be controlled by gating biases. Therefore, the electrostatic gating can not affect the emission current under high across-gap voltage, but affect the emission current under low across-gap bias. These analyses are consistent with the experimental results we observed. Actually, the electrical field distribution near SLG edges is more complicated in real cases. Firstly, electrons in SLG behave not like the free electrons in metal. As the density of states in graphene is low near the $K$ point, SLG may be subjected to band-bending under high electrical field.[23] In addition, there are electronic states at graphene edges, where SLG is subject to electron localization. For these reasons, SLG may exhibit both internal voltage drops near edges and statistical decoupling of the edge-state electron distribution from the electron distribution in graphene body. Therefore, the electrical field distribution near edges may become very complex in real cases.

In summary, we demonstrated a technique by which SLG nanogap can be achieved with crystallographically parallel edges. We investigated I-V characteristics of a lateral graphene nanogap inside a vacuum chamber. From the I-V curves, the transition from the classic Child-Langmuir law at low current region to the Fowler-Nordheim law at high current region is observed in the graphene nanogap devices. Electrons emitted from SLG graphene are almost independent of the gate voltage within the range from -80V to +80V. This research could enable the construction of graphene based vacuum nanoelectronics with many far-reaching potential applications.

**Figure Caption:**

**Figure 1** Schematic diagram of fabrication processes. (a) Graphene flakes are transferred onto the heavily doped silicon substrate coated with 300nm $SiO_2$ and 100nm PMMA. (b) Covering the sample with another thick layer of PMMA (about 300nm). (c) Exposure to e-beam and developing in MIBK solutions open windows for electrodes. (d) Evaporation of metal contacts. A strong metal bridge it possible when the total electrode thickness is comparable to bottom PMMA thickness. Metals are evaporated at 45 degree onto the substrate surface. e) Resists are lifted off in PG remover. A free standing graphene flake forms after lift-off. (f) The suspended graphene collapses and is torn apart by the surface tension of solvent during the drying process.

**Figure 2** (a) An optical image of SLG on PMMA(100nm)/SiO$_2$(300nm)/Si. The outline area corresponds to SLG, the scale bar is 20μm. (b) Experimental results of contrast spectra of the graphene sample. (c) Raman spectrum of the monolayer graphene flake obtained with WITEC CRM200 Raman system using a 532 nm Ar line as an excitation source. (d) Scanning electron micrograph of a representative SLG nanogap with suspended electrodes taken at 60 degree. The arrows point out the edges of graphene flakes.

**Figure 3** (a) Room temperature current-voltage characteristics of the individual graphene nanogap shown in Fig. 2(d) under different gate bias. (b) The corresponding Fowler–Nordheim ($\ln(I/V^2)$ vs $V^{-1}$) plots. (c) The plot of lnI vs lnV curve obtained for the graphene sample. Voltage is measured in volts while current is measured in $10^{-10}$ A. (d) The gate-voltage dependence of field enhancement factor β in Fowler–Nordheim law and the power α in the Child-Langmuir law.

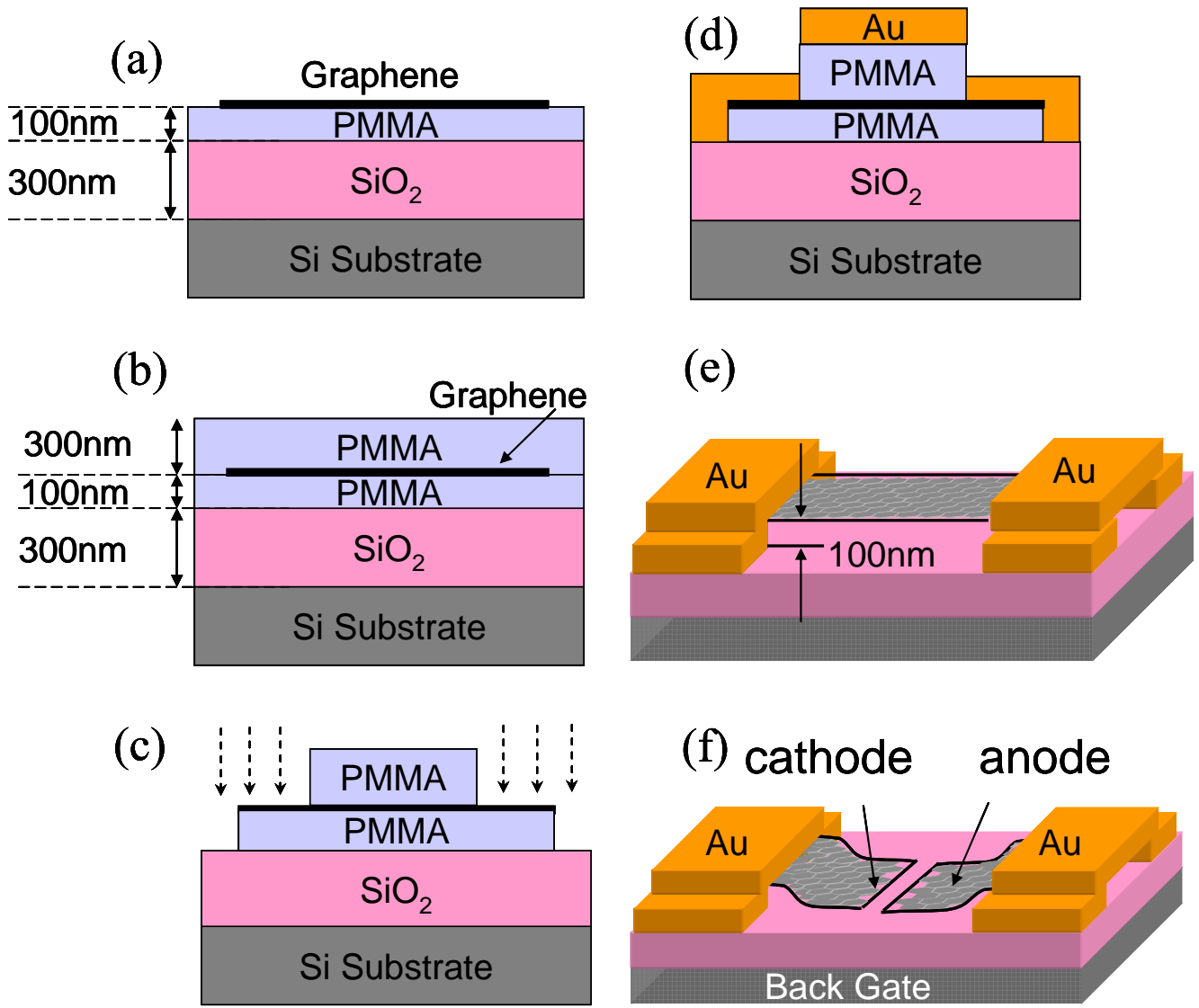

Fig.1 Wang et al, APL

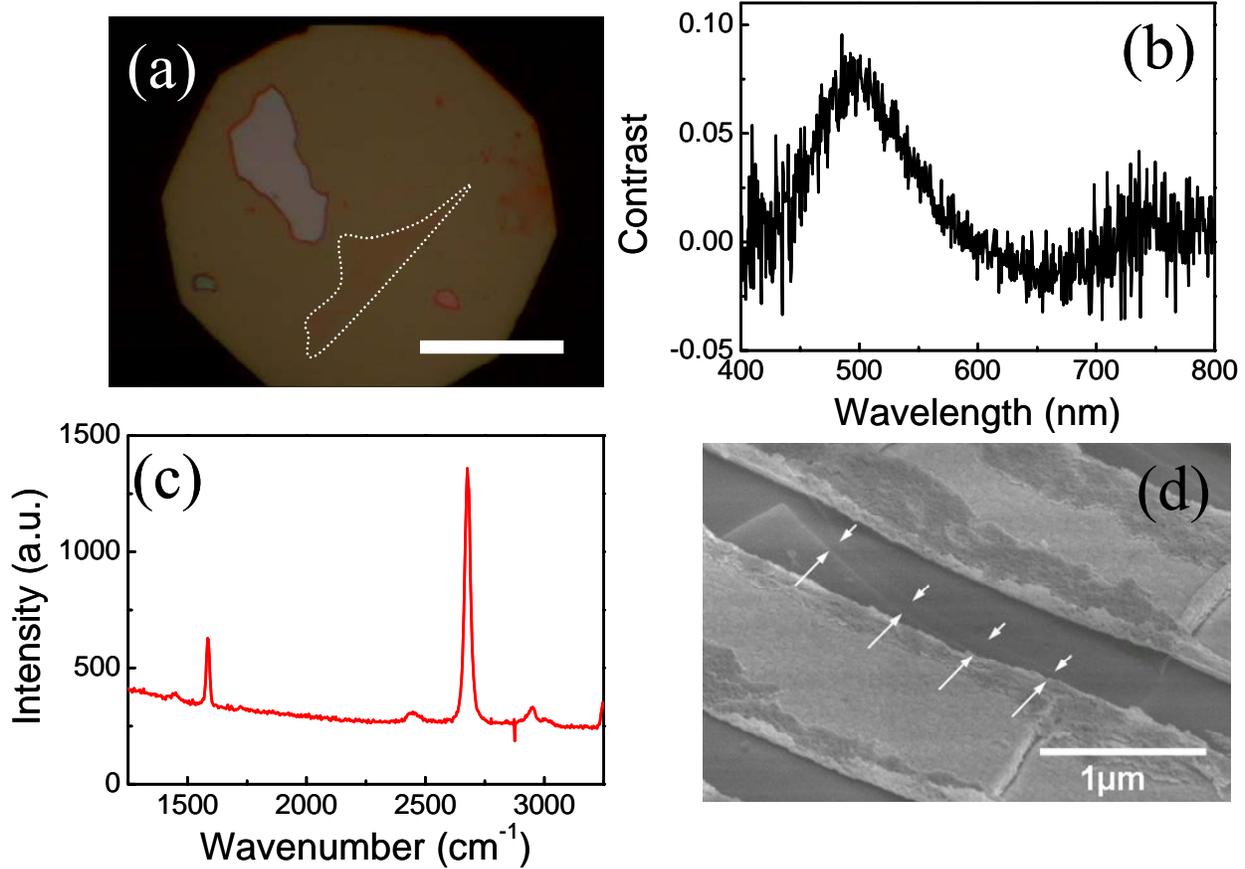

Fig.2 Wang et al, APL

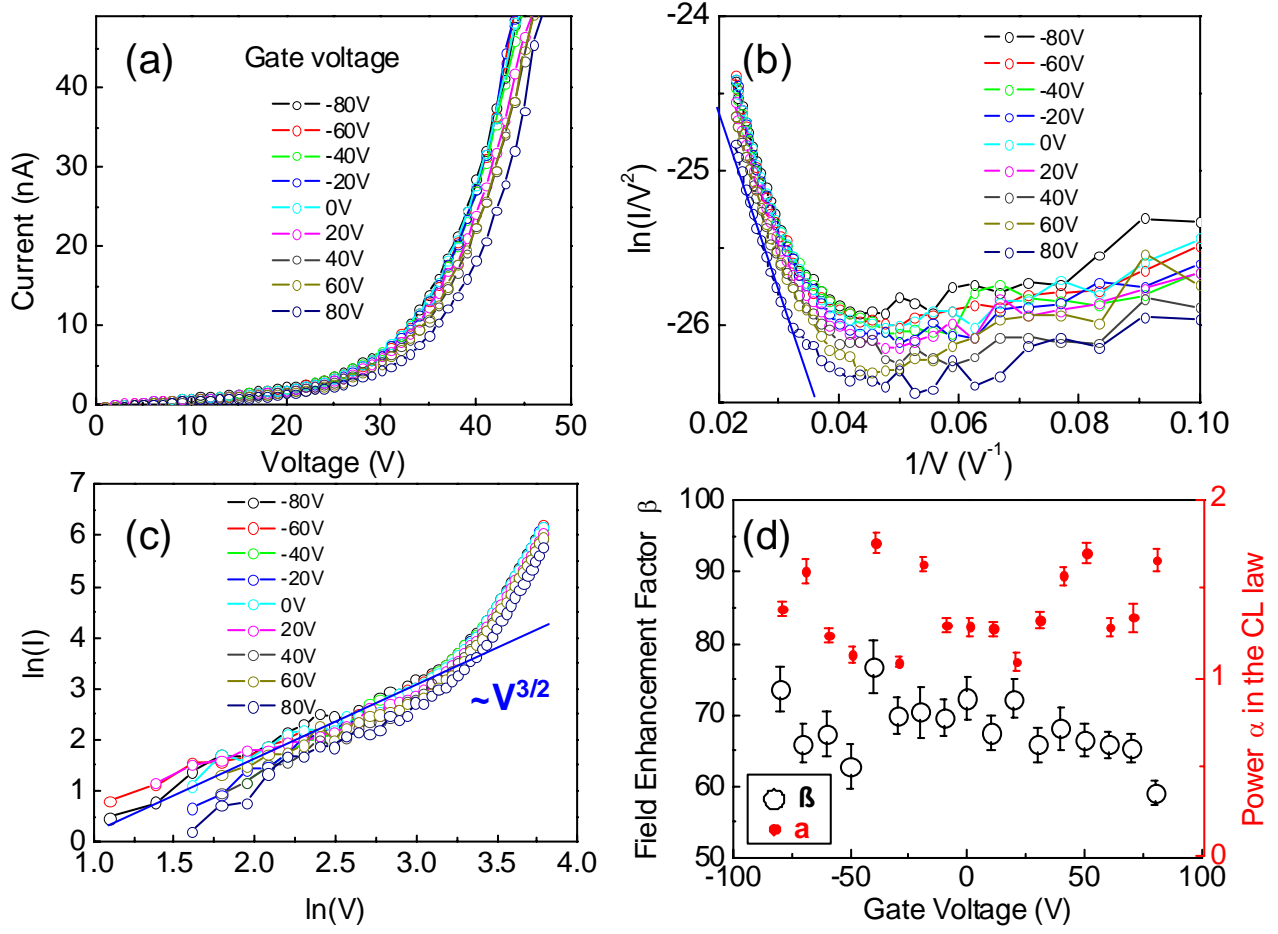

Fig.3 Wang et al, APL